\documentclass[aps,showpacs,preprintnumbers,nofootinbib]{revtex4}
\usepackage{amssymb}
\usepackage{amsmath}
\usepackage{bm}

\input{tcilatex}

\begin{document}

\title{Inducing the cosmological constant from five-dimensional Weyl space }
\author{Jos\'{e} Edgar Madriz Aguilar\thanks{%
E-mail address: jemadriz@fisica.ufpb.br}, and Carlos Romero \thanks{%
E-mail address: cromero@fisica.ufpb.br} }
\affiliation{$^{a}$Departamento de F\'{\i}sica, Universidade Federal da Para\'{\i}ba,
Caixa Postal 5008, 58059-970 \\
Jo\~{a}o Pessoa, PB, Brazil\\
E-mail: jemadriz@fisica.ufpb.br; cromero@fisica.ufpb.br}

\begin{abstract}
We investigate the possibility of inducing the cosmological constant from
extra dimensions by embedding our four-dimensional Riemannian space-time
into a five-dimensional Weyl\ integrable space. Following approach of the
induced matter theory \ we show that when we go down from five to four
dimensions, the Weyl field may contribute both to the induced energy-tensor
as well as to the cosmological constant $\Lambda $, or more generally, it
may generate a time-dependent cosmological parameter $\Lambda (t).$ As an
application, we construct a simple cosmological model in which $\Lambda (t)$
has some interesting properties.
\end{abstract}

\maketitle

\vskip .5cm


\section{Introduction}

In a very recent past it appeared that the role played by the cosmological
constant in Cosmology was merely historical, mainly connected with
Einstein's attempt to build a cosmological scenario in which the Universe
was static and finite \cite{Einstein} . However, since the recent discovery
of cosmic acceleration there has been a renewed interest in the role the
cosmological constant could play to explain the new data. For instance,
there is strong evidence that the so-called dark energy, which would cause
the Universe's accelerated expansion, might have a connection with the
cosmological constant. Moreover, the present most popular model of
cosmology, the Lambda-CDM model, tacitally assumes the existence of the
cosmological constant \cite{tegmark}. On the other hand, particle physics
theorists have always argued if favour of the existence of the cosmological
constant as a consequence of the energy density of the vacuum \cite{Weinberg}%
.

From the standpoint of cosmological theory it seems then desirable to have a
justification of the cosmological constant on theoretical grounds. This
quest has led some theoreticians to modify Einstein's gravitational theory,
these attempts going back to the works of Eddington and Schr\"{o}dinger \cite%
{Goenner}.

Our aim in the present article is to introduce a new approach to this old
question. We argue that the appearance of the cosmological constant in the
equations of four-dimensional (4D) general relativity might be related to
the assumption, made by some modern spacetime theories, that our Universe
may have extra dimensions. Among such theories are the so-called braneworld
scenario \cite{Arkani} and the non-compact Kaluza-Klein theories \cite{ndGR}%
, both sharing some basic assumptions concerning the geometry of the
fundamental higher-dimensional embedding space. For example, in these
proposals our ordinary spacetime is viewed as a hypersurface (\textit{the
brane}) embedded in a five-dimensional (5D) manifold (\textit{the bulk}). On
the other hand, mathematical theorems regulate these embeddings, in
particular, the Campbell-Magaard theorem \cite{campbell} and its extensions
specify the conditions under which the embeddings are possible \cite{Tavakol}%
.

The possibility of generating matter and fields from a higher-dimensional
vacuum was first realized by T. Kaluza and O. Klein, an idea that has been a
source of inspiration for practically all higher-dimensional theories \cite%
{kaluza}. In particular, a mathematical formalism has been developed by
Wesson and Ponce de Leon \cite{Ponce} that permits to "induce" an
energy-momentum tensor from the vacuum Einstein field equations in five
dimensions, a schema now known as the induced-matter proposal. It has later
been shown that from a geometrical point of view such generated
energy-momentum tensor is nothing more than the extrinsic curvature of the
spacetime hypersurface in disguise \cite{maia} . Besides, an interesting
feature of this mechanism, apart from this "geometrization" of matter, is
that it is also powerful enough to generate a cosmological constant in four
dimensions out of a pure five-dimensional vacuum \cite{Ponce2}. It should be
noted, however, that in the context of the usual induced matter approach by
considering a Riemannian bulk it has not been possible until now\ to
generate simultaneously the cosmological constant and an induced
energy-momentum tensor $T_{\alpha \beta }^{(IM)}$ that describes macroscopic
matter, at least without modifying the "canonical" form of $T_{\alpha \beta
}^{(IM)}$ \cite{ndGR} . The interesting point here is that this apparent
weakness can be remedied if we allow for the geometry of the embedding space
having more degrees of freedom such as in the case of Weyl geometry \cite%
{Weyl}. It turns out then that when we go down from 5D to 4D, as in the
induced matter approach, the Weyl field may contribute both to the induced
energy-tensor as well as to the cosmological constant $\Lambda $, or more
generally, to a time-dependent cosmological parameter $\Lambda (t)$.

In this paper we shall consider a particular case of Weyl geometry, namely,
the one in which the Weyl field is integrable \cite{Pauli}. It turns out
that in this case if the Weyl field depends only on the extra dimension,
then the embedded spacetime is Riemannian and general relativity holds in
the brane \cite{we1}, although the non-Riemannian character of the whole
bulk propagates into the brane in the form of induced matter and a
cosmological parameter.

The paper is organized as follows. We start in Section I with a brief review
of some fundamental concepts that underlie Weyl geometry. We proceed in
Section II to develop a\ \ five-dimensional Weylian theory of gravity in
vacuum, the dynamics of which is given by a certain action chosen among the
simplest ones \cite{Novello}. In the same section we set the equations for a
particular choice of the five-dimensional metric and examine the case of a
simple cosmological model. In Section III we illustrate the theory by going
through a simple example taken from Cosmology. Our final remarks are
contained in Section IV.

\section{Weyl geometry}

Conceived by H. Weyl in 1918, as an attempt to unify gravity with
electromagnetism, in its original form Weyl%
\'{}%
s theory \cite{Weyl} turned out to be innadequate as a physical theory as
was firstly pointed by Einstein soon after the appearance of the theory \cite%
{Pauli}. However, a variant of Weyl geometry, known as \textit{Weyl
integrable geometry} does not suffer from the drawback pointed out by
Einstein and for this reason has attracted the attention of some
cosmologists \cite{Novello}, particularly in the context of
higher-dimensional theories \cite{Israelit}.

The starting point of the new geometry created by Weyl is the assumption
that the covariant derivative of the metric tensor $g$ is not zero, but,
instead, is given by

\begin{equation}
\nabla _{a}g_{bc}=\sigma _{a}g_{bc}  \label{compatibility}
\end{equation}%
where $\sigma _{a}$ \ denotes the components of a one-form field $\sigma $
with respect to a local coordinate basis. This is, of course, a
generalization of the idea of Riemannian compatibility between the
connection $\nabla $ and $g,$ which is equivalent to require the length of a
vector to remain unaltered by parallel transport \cite{Pauli}. If $\sigma $ $%
=d\phi ,$ where $\phi $ is a scalar field, then we have an integrable Weyl
geometry. A differentiable manifold $M$ endowed with a metric $g$ and a Weyl
field $\sigma $ $\ $is usually referred to as a \textit{Weyl frame}. It is
interesting to see that the Weyl condition (\ref{compatibility}) remains
unchanged\ when we go to another Weyl frame $(M,\overline{g},\overline{%
\sigma })$ by performing the following simultaneous transformations in $g$
and $\sigma $:%
\begin{equation}
\overline{g}=e^{-f}g  \label{conformal}
\end{equation}%
\begin{equation}
\overline{\sigma }=\sigma -df  \label{gauge}
\end{equation}%
where $f$ is a scalar function defined on $M$.

Our next step is to consider a Weylian theory of gravity. If we consider a
Weyl spacetime the simplest action that gives the dynamics of the
gravitational field in the absence of matter is given by 
\begin{equation}
\mathcal{S}=\int d^{4}x\,\sqrt{g}\left[ \!\mathcal{R}+\xi \phi ^{a}\,_{;a}%
\right]  \label{weyldynamics}
\end{equation}%
where $\xi $ is an arbitrary coupling constant, $\phi _{a}\equiv \phi _{,a}$
is the the Weyl field, $\!\mathcal{R}$ is the Weylian Ricci scalar and the
semicolon $(;)$ denotes covariant derivative with respect to the Weyl
connection \cite{Novello}. The extension of this formulation to a
higher-dimensional space is straightforward. In the next section we shall
consider the a five-dimensional Weyl integrable space.

\section{A Weyl integrable dynamics in five dimensions}

Let us consider a five-dimensional space $M^{5}$ endowed with a metric
tensor $^{(5)}g$ and an integrable Weyl scalar field $\phi $. In local
coordinates $\{y^{a}$$\}$ the five-dimensional line element of \ will be
denoted by 
\begin{equation}
dS^{2}=g_{ab}(y)\,dy^{a}dy^{b},  \label{w5d1}
\end{equation}

where \ $g_{ab}$ are the components of $^{(5)}g.$\ As we already mentioned
the simplest action that can be constructed for a Weylian theory of gravity
in \ a five-dimensional vacuum is given by 
\begin{equation}
^{(5)}\!\mathcal{S}=\int d^{5}y\,\sqrt{\left\vert ^{(5)}g\right\vert }\left[
^{(5)}\!\mathcal{R}+\xi \phi ^{a}\,_{;a}\right] ,  \label{w5d2}
\end{equation}%
where $\xi $ is an arbitrary coupling constant, $\phi _{a}\equiv \phi _{,a}$
is the gauge vector associated to the Weyl field, $\left\vert
^{(5)}g\right\vert $ is the absolute value of \ the determinant of the
metric $^{(5)}g_{ab}$, $^{(5)}\!\mathcal{R}$ is the Weylian Ricci scalar.
One can easily check that the variation of the action (\ref{w5d2}) with
respect to the tensor metric and with respect to the Weyl scalar field
yields 
\begin{equation}
^{(5)}\!\mathcal{G}_{ab}+\phi _{a;b}-(2\xi -1)\phi _{a}\phi _{b}+\xi
g_{ab}\phi _{c}\phi ^{c}=0,  \label{w3}
\end{equation}%
\begin{equation}
\phi ^{a}\,_{;a}+2\,\phi _{a}\phi ^{a}=0  \label{w4}
\end{equation}%
where $^{(5)}\!\mathcal{G}_{ab}$ denotes the Einstein tensor calculated with
the Weyl connection $^{(5)}\Gamma _{bc}^{a}=^{(5)}\{_{bc}^{\,a}\}-(1/2)[\phi
_{b}\delta _{c}^{a}+\phi _{c}\delta _{b}^{a}-g_{bc}\phi ^{a}]$ and $%
\{_{bc}^{\,a}\}$ are the Christoffel symbols of Riemannian geometry. The
equations (\ref{w3}) and \ (\ref{w4}) are the field equations of the
five-dimensional Weyl gravitational theory and describes the dynamics of a
five-dimensional bulk in vacuum. A better insight may be gained if we recast
the field equations (\ref{w3}) and (\ref{w4}) into its Riemannian part plus
the contribution of the Weyl scalar field. Thus after excluding total
derivatives of the scalar field $\phi $ the action (\ref{w5d2}) can be
written as \cite{Novello} 
\begin{equation}
^{(5)}\mathcal{S}=\int d^{5}y\,\sqrt{g_{5}}\left[ ^{(5)}\!\tilde{R}+\frac{1}{%
2}(5\xi -6)\phi _{a}\phi ^{a}\right]  \label{nueva5D1}
\end{equation}%
The field equations are obtained by taking a variation of the above action
with respect to the pair $(g_{ab},\phi )$. We are then led to 
\begin{eqnarray}
&&^{(5)}\tilde{G}_{ab}-\frac{1}{2}(6-5\xi )\left[ \phi _{a}\phi _{b}-\frac{1%
}{2}g_{ab}\phi _{c}\phi ^{c}\right] =0,  \label{nueva5D3} \\
&&^{(5)}\tilde{\Box}\phi =0,
\end{eqnarray}%
where the tilde $(\sim )$ is used to denote quantities calculated with the
Riemannian part of the Weyl connection and $^{(5)}\tilde{\Box}$ denotes the
5D d'Alembertian operator in the Riemannian sense.\newline

At this point let us express the local coordinates $\{y^{a}\}$ as $%
\{x^{\alpha },l\}$ denoting \ by $l$ the fifth (spacelike) coordinate and we
choose for simplicity the line element (\ref{w5d1}) in the form \footnote{%
We shall adopt the convention $diag(+---)$ for the signature of $g_{\alpha
\beta }.$} 
\begin{equation}
dS^{2}=g_{\alpha \beta }(x,l)dx^{\alpha }dx^{\beta }-\Phi ^{2}(x,l)dl^{2}.
\label{metric}
\end{equation}%
where the function $\Phi ^{2}(x,l)$ is the 5D analogue of the lapse function
used in canonical general relativity, which supposes that spacetime may be
foliated by a family of spacelike surfaces \cite{Wheeler}.

As in the induced matter approach \cite{Ponce}, with respect to this
foliation the field equations (\ref{nueva5D3}) can 
be splitted as

\begin{equation}
^{(5)}\tilde{G}_{\alpha \beta }-\frac{1}{2}(6-5\xi )\left[ \phi _{\alpha
}\phi _{\beta }-\frac{1}{2}g_{\alpha \beta }\left( \phi _{\gamma }\phi
^{\gamma }-\Phi ^{-2}\phi _{l}^{2}\right) \right] =0,  \label{nueva5D6a}
\end{equation}%
\begin{equation}
^{(5)}\tilde{G}_{\alpha l}-\frac{1}{2}(6-5\xi )\phi _{\alpha }\phi _{l}=0,
\label{nueva5D7a}
\end{equation}%
\begin{equation}
^{(5)}\tilde{G}_{ll}-\frac{1}{4}(6-5\xi )\left[ \phi _{l}^{2}+\Phi ^{2}\phi
_{\gamma }\phi ^{\gamma }\right] =0.  \label{nueva5D8a}
\end{equation}
The \ five-dimensional Weylian equations we have now obtained assuming the
geometry given by (\ref{metric}) supposes that in principle the Weyl scalar
field $\phi $\ depends on all coordinates, that is, $\phi =\phi (x,l)$. Some
solutions of the above field equations have been worked out in detail by
Novello and colaborators considering different geometric settings \cite%
{Novello}. However our interest here is mainly to study a particular case of
these field equations when the four-dimensional spacetime can be embedded in
a five-dimensional ambient space whose dynamics comes from an integrable
Weyl theory of gravity. In fact this is strongly motivated by a recently
result concerning the existence of\ necessary and sufficient conditions for
a Riemannian manifold to be embedded in a Weyl space \cite{we1}. According
to these, if the Weyl scalar field depends only on the extra coordinate $l$,
then each leaf of the foliation $l=const$ has a Riemannian character and can
be locally and isometrically embedded in a five-dimensional Weylian space
whose metrical properties are given by (\ref{metric}). Since we regard the
spacetime as one of the leaves of the foliation and given that such
embedding preserves the Riemannian character of the spacetime we proceed to
investigate the four-dimensional field dynamics induced by the
five-dimensional space. In much the same way as in induced-matter theory 
\cite{Ponce} one would interpret the extra contributions coming from the
extra dimension as macrosocopic matter in 4D.

In view of the above let us assume that $\phi =\phi (l)$, i.e\ the Weyl
scalar field depends only on the extra coordinate $l$. In this case the
field equations (\ref{nueva5D6a}), (\ref{nueva5D7a}) and (\ref{nueva5D8a})
become 
\begin{eqnarray}
&&^{(5)}\tilde{G}_{\alpha \beta }+\frac{1}{4}(5\xi -6)\Phi ^{-2}g_{\alpha
\beta }\phi _{l}^{2}=0,  \label{nueva5D7} \\
&&^{(5)}\tilde{G}_{\alpha l}=0,  \label{nueva5D8} \\
&&^{(5)}\tilde{G}_{ll}-\frac{1}{4}(6-5\xi )\phi _{l}^{2}=0,  \label{nueva5D9}
\\
&&\frac{\partial }{\partial l}\left[ \sqrt{\left\vert g_{5}\right\vert }%
\,\Phi ^{-2}\phi _{l}^{2}\right] =0.  \label{nueva5D10}
\end{eqnarray}%
To illustrate with an example let the five-dimensional space $M^{5}$
correspond to a five-dimensional cosmological model in the form of a triple
warped product manifold \cite{Dahia1} with metric given by 
\begin{equation}
dS^{2}=dt^{2}-a^{2}(t)dr^{2}-e^{2F(t)}dl^{2},  \label{nueva5D11}
\end{equation}%
where $dr^{2}=\delta _{ij}dx^{i}dx^{j}$ is the three-dimensional Euclidian
line element, $t$ represents the cosmic time for co-moving observers, $F(t)$
is a well-behaved real function and $a(t)$ is the cosmological scale factor.
Inserting the metric (\ref{nueva5D11}) in (\ref{nueva5D10}) it can be easily
seen that the Weyl scalar field in this case is given by 
\begin{equation}
\phi (l)=C_{1}l+C_{2},  \label{nueva5D12}
\end{equation}%
where $C_{1}$ and $C_{2}$ are integration constants. The field equations (%
\ref{nueva5D7}), (\ref{nueva5D8}) and (\ref{nueva5D9}) now give 
\begin{eqnarray}
3H^{2}+3\dot{F}H &=&\frac{1}{4}(6-5\xi )C_{1}^{2}e^{-2F},  \label{nueva5D13}
\\
2\frac{\ddot{a}}{a}+H^{2}+2\dot{F}H+\ddot{F}+\dot{F}^{2} &=&\frac{1}{4}%
(6-5\xi )C_{1}^{2}e^{-2F},  \label{nueva5D14} \\
3\left( \frac{\ddot{a}}{a}+H^{2}\right) &=&-\frac{1}{4}(6-5\xi
)C_{1}^{2}e^{-2F},  \label{nueva5D15}
\end{eqnarray}%
where $H(t)=\dot{a}/a$ is the Hubble parameter. From (\ref{nueva5D14}) and (%
\ref{nueva5D15}) we obtain the equation 
\begin{equation}
\ddot{F}+\dot{F}^{2}+2H\dot{F}+5\frac{\ddot{a}}{a}+4H^{2}=0.
\label{nueva5D16}
\end{equation}%
In order to simplify the structure of this equation we introduce a new
function $u(t)$ defined by $u(t)=a(t)e^{F(t)}$. In this way (\ref{nueva5D16}%
) becomes 
\begin{equation}
\ddot{u}+4\left( \frac{\ddot{a}}{a}+H^{2}\right) u=0.  \label{nueva5D17}
\end{equation}%
The above equation relates $u(t)$ with $a(t)$ in such a way that the
solutions of (\ref{nueva5D17}) can be substituted in (\ref{nueva5D13})
yielding a differential equation for $a(t)$, which in principle can be
solved.

\section{The dynamics induced on the four-dimensional Riemannian brane}

As we have mentioned in the previous section one of our aims is to explore
the possibility of interpreting the extra contributions of the
five-dimensional Weylian bulk to the four-dimensional Riemannian brane as
four-dimensional matter induced geometrically. In this section we shall
study the four-dimensional dynamics geometrically induced on a generic
brane. We recalll that we are assuming that the five-dimensional space is
foliated by a family of hypersurfaces $\{\Sigma \}$ defined by the equation $%
l=const$. Clearly, on a particular hypersurface $\Sigma _{0}$ the induced
line element will be given by 
\begin{equation}
dS_{\Sigma _{0}}^{2}=h_{\alpha \beta }(x)dx^{\alpha }dx^{\beta },
\label{w4d1}
\end{equation}%
where $h_{\alpha \beta }(x)=g_{\alpha \beta }(x,l_{0})$ is the induced
metric on $\Sigma _{0}$. From the Gauss-Codazzi equations it is easy to show
(See, for instance \cite{Israelit}) that the induced dynamics on the
hypersurface $\Sigma _{0}$ is governed by the four-dimensional field
equations 
\begin{equation}
^{(4)}\tilde{G}_{\alpha \beta }=T_{\alpha \beta }^{(IM)}+\Lambda
(x)h_{\alpha \beta },  \label{nu4d1}
\end{equation}%
where $T_{\alpha \beta }^{(IM)}$ is the usual energy momentum tensor
obtained in the induced matter approach, which has the form \cite{Ponce} 
\begin{equation}
T_{\alpha \beta }^{(IM)}=\frac{\Phi _{\alpha ||\beta }}{\Phi }+\frac{1}{%
2\Phi ^{2}}\left\{ \frac{\overset{\star }{\Phi }}{\Phi }\overset{\star }{g}%
_{\alpha \beta }-\overset{\star \star }{g}_{\alpha \beta }+g^{\lambda \mu }%
\overset{\star }{g}_{\alpha \lambda }\overset{\star }{g}_{\beta \mu }-\frac{1%
}{2}g^{\mu \nu }\overset{\star }{g}_{\mu \nu }\overset{\star }{g}_{\alpha
\beta }+\frac{1}{4}g_{\alpha \beta }\left[ \overset{\star }{g}^{\mu \nu }%
\overset{\star }{g}_{\mu \nu }+\left( g^{\mu \nu }\overset{\star }{g}_{\mu
\nu }\right) ^{2}\right] \right\} ,  \label{nu4d2}
\end{equation}%
with the bars $(||)$ denoting covariant derivative in a Riemannian sense and
the star $(\star )$ denoting derivative with respect to the fifth coordinate 
$l$, and the function $\Lambda (x)$ is given by 
\begin{equation}
\Lambda (x)=\frac{1}{4}(6-5\xi )\Phi ^{-2}\left. \phi _{l}^{2}\right\vert
_{l=l_{0}}.  \label{nu4d3}
\end{equation}%
Clearly, both terms $T_{\alpha \beta }^{(IM)}$ and $\Lambda (x)$ comes from
the Weylian bulk. The induced energy-momentum tensor $T_{\alpha \beta
}^{(IM)}$\ can be obtained even if the bulk is Riemannian, but the
interesting fact here is that the function $\Lambda (x)$ is a new
contribution depending directly on the Weyl scalar field. It is worth
mentioning that when the lapse function $\Phi $ dependends only on the time
then $\Lambda (t)$ can be interpreted as an induced cosmological parameter,
whereas if $\Phi $ is constant then \ref{nu4d4} \ reduces to an induced
cosmological constant.

\section{A simple aplication to Cosmology}

As a simple application of the ideas developed in the previous section let
us have a quick look into the cosmological scenario that takes place in the
four-dimensional brane $\Sigma _{0},$ whose geometry is induced by the line
element (\ref{nueva5D11}). In this case the induced line element (\ref{w4d1}%
) becomes 
\begin{equation}
dS_{\Sigma _{0}}^{2}=dt^{2}-a^{2}(t)dr^{2},  \label{nu4d4}
\end{equation}%
which is nothing more than the line element of a Friedmann-Robetson-Walker
model. The induced energy-momentum tensor (\ref{nu4d2}) reduces to 
\begin{equation}
T_{\alpha \beta }^{(IM)}=F_{,\alpha ,\beta }+F_{,\alpha }F_{,\beta
}-\,^{(4)}\!\{\,_{\alpha \beta }^{\gamma }\}F_{,\gamma }\,\,,  \label{nu4d5}
\end{equation}%
where the $^{(4)}\!\{\,_{\alpha \beta }^{\gamma }\}$ denote the
four-dimensional Christoffel symbols calculated with the induced metric $%
h_{\mu \nu }$ in (\ref{nu4d4}). Assuming that the induced matter
configuration\ given by (\ref{nu4d5}) is that of a perfect fluid, as viewed
by four-dimensional comoving observers located at the brane $\Sigma _{0}$,
we can define the energy density $\rho _{(IM)}$ and pressure $P_{(IM)}$ for
the induced matter by $\rho _{(IM)}=T^{(IM)}\,^{t}\,_{t}$ and $%
P_{(IM)}=-T^{(IM)}\,^{r}\,_{r}$ respectively. \ Thus using (\ref{nu4d4}) the
4D field equations (\ref{nu4d1}) becomes 
\begin{eqnarray}
3H^{2} &=&\rho _{(IM)}+\Lambda (t),  \label{nu4d6} \\
2\frac{\ddot{a}}{a}+H^{2} &=&-\left( P_{(IM)}-\Lambda (t)\right) ,
\label{nu4d7}
\end{eqnarray}%
where according to (\ref{nu4d3}) and (\ref{nueva5D12}) the induced varying
cosmological \textquotedblleft constant" $\Lambda (t)$ is given by 
\begin{equation}
\Lambda (t)=\left( \frac{C_{1}}{2}\right) ^{2}(6-5\xi )e^{-2F(t)}.
\label{nu4d8}
\end{equation}%
Introducing the effective energy density $\rho _{eff}=\rho _{(IM)}+\Lambda
(t)$ and the effective pressure $P_{eff}=P_{(IM)}-\Lambda (t)$ we define a
parameter $\omega _{eff}$ associated with the effective equation of state,
which is given by 
\begin{equation}
\omega _{eff}\equiv \frac{P_{eff}}{\rho _{eff}}=-\left[ 1-\frac{\dot{F}^{2}+%
\ddot{F}-H\dot{F}}{\ddot{F}+\dot{F}^{2}+\Lambda (t)}\right] .  \label{nu4d9}
\end{equation}%
By simple inspection it can easily be seen that $\omega _{eff}$ depends
entirely on the metric function $F(t)$, which, in turn, can be determined by
the bulk dynamics, i.e. by finding solutions of the system (\ref{nueva5D13}%
)-(\ref{nueva5D15}). A particular solution $F(t)$ for a given scale factor $%
a(t)$ can be obtained by solving the equation (\ref{nueva5D17}). Thus if we
look for solutions $F(t)$ in the case of a power-law expanding universe with
the scale factor given by $a(t)=a_{0}(t/t_{0})^{p}$, the equation (\ref%
{nueva5D17}) becomes 
\begin{equation}
\ddot{u}+\frac{4p(2p-1)}{t^{2}}u=0,  \label{nu4d10}
\end{equation}%
whose general solution is given by 
\begin{equation}
u(t)=A_{1}t^{1/2+(1/2)\sqrt{1-32p^{2}+16p}}+A_{2}t^{1/2-(1/2)\sqrt{%
1-32p^{2}+16p}},  \label{nu4d11}
\end{equation}%
where $A_{1}$ and $A_{2}$ \ are integration constants. Moreover, choosing $%
A_{2}=0$, the corresponding particular solution for $F(t)$ can be written as 
\begin{equation}
F(t)=ln(B_{1}t^{\gamma }),  \label{nu4d12}
\end{equation}%
where $B_{1}=(A_{1}t_{0}^{p}/a_{0})$ and $\gamma =(1/2-p)+(1/2)\sqrt{%
1-32p^{2}+16p}$\thinspace . Note that if we want to have real values for the
power $\gamma $ that are compatible with an expanding universe ($p>0$), the
values of $p$ must range in the interval $0<p\leq (1/4)+\sqrt{6}/8$. On the
other hand, \ if we insert (\ref{nu4d12}) into (\ref{nu4d8}) and (\ref{nu4d9}%
), then the induced variable cosmological \textquotedblleft constant" $%
\Lambda (t)$ and the effective parameter $\omega _{eff}$ are given,
respectively, by 
\begin{eqnarray}
\Lambda (t) &=&\left( \frac{C_{1}}{2}\right) ^{2}(6-5\xi
)B_{1}^{-2}t^{-2\gamma },  \label{nu4d13} \\
\omega _{eff} &=&-\left[ 1-\frac{\gamma ^{2}-\gamma -p\gamma }{\gamma
^{2}-\gamma +(C_{1}/2)^{2}(6-5\xi )B_{1}^{-2}t^{2-2\gamma }}\right] .
\label{nu4d14}
\end{eqnarray}%
If we want to have $\omega _{eff}$\ decreasing with time we must require $%
2-2\gamma >0$, and this condition restricts the range of variation of the
parameter $p$ to $p>1/3$. Finally, if the former inequality is to be
compatible with an expanding universe $p$ must range in the interval $%
1/3<p\leq (1/4)+\sqrt{6}/8$. One reason for restricting the parameter $p$ to
this interval is that with a suitable choice of $p$ the effective parameter $%
\omega _{eff}$ will tend asymptotically to $-1$. We conclude that, in its
final state, our model would tend to a de Sitter universe. Finally, note
that in the case when $p=5/9$ the induced $\Lambda (t)$ given by (\ref%
{nu4d13}) becomes a constant, while the value of $\omega _{eff}$ \ is
exactly $-1$. This means that, in such models, $p=5/9$ corresponds to a de
Sitter universe.

\section{Final remarks}

In this paper we have considered the idea of generating a cosmological
constant, or rather, a cosmological parameter, from extra dimensions.
Although this has already been investigated in the context of induced matter
theory, the novelty of our approach is to regard the same problem in\ a more
general setting, i.e by assuming the geometry of the embedding space to have
a Weylian character. Two comments are in order: Firstly, the embedding space
has a prescribed dynamics; secondly, the embedding does not affect the
Riemannian geometry of the spacetime. These features depend on the fact that
theWeyl field is assumed to be integrable and depending only on the extra
dimension. Finally, by setting up a simple\ "toy model " our intention is to
call attention to the richness of non-Riemannian geometries, in particular
to the Weyl integrable manifolds, as a way of providing new degrees of
freedom that might play a role in the theoretical framework of
higher-dimensional embedding theories of spacetime. We believe that in this
context issues such as the nature of the cosmological constant, dark energy
and other important questions may be investigated from an entirely new point
of view.

\section*{Acknowledgements}

\noindent The authors would like to thank CNPq-CLAF and CNPq-FAPESQ (PRONEX)
for financial support. We are indebted to Dr. F. Dahia for helpful comments.


\bigskip

\begin{thebibliography}{99}
\bibitem{Einstein} A. Einstein, A., \textquotedblleft Zum kosmologischen
Problem der allgemeinen Relativit\"{a}tstheorie\textquotedblright , Sitz.
Ber.$\backslash $ Preuss. Akad. Wiss., 142, 235-237, (1931).

\bibitem{tegmark} See, for instance, M. Tegmark et al.,Phys. Rev. D69 103501
(2004).

\bibitem{Weinberg} S. Weinberg, Rev. Mod. Phys., 61, 1 (1989).

\bibitem{Goenner} See, for instance, H. F. M. Goenner, "On the History of
Unified Field Theories", \textit{Living Rev. Rel}., \textbf{7}, 2 (2004)

\bibitem{Arkani} N. Arkani-Hamed, S. Dimopoulos, and G. Dvali, Phys. Lett. B 
\textbf{429}, 263 (1998); I. Antoniadis, N. Arkani-Hamed, S. Dimopoulos, and
G. Dvali, Phys. Lett. B \textbf{436}, 257 (1998). L. Randall and R. Sundrum,
Phys. Rev. Lett. 83, 3370 (1999); L. Randall and R. Sundrum, Phys. Rev.
Lett. 83, 4690 (1999).

\bibitem{ndGR} P.S. Wesson and J. Ponce de Leon, J. Math. Phys. 33, 3883
(1992); J.M. Overduin and P.S. Wesson, Phys. Rept. 283 ,303 (1997),
gr-qc/9805018; P.S. Wesson, \textit{Space-Time-Matter}, (World Scientific,
Singapore 1999); P. S. Wesson, \textit{Five-Dimensional Physics} (World
Scientific, 2006).

\bibitem{campbell} J. E. Campbell, \textit{A Course of Differential Geometry}%
, Clarendon, 1926. L. Magaard, \textit{Zur Einbettung Riemannscher Raume in
Einstein-raume und konform-euclidische Raume}, PhD Thesis, Kiel, 1963

\bibitem{Tavakol} C. Romero, R. Tavakol, and R. Zalaletdinov, Gen. Rel.
Grav. \textbf{28}, 365 (1995). F. Dahia and C. Romero, J. Math. Phys. 
\textbf{43}, 5804 (2002). E. Anderson and J. E. Lidsey, Class. Quant. Grav.%
\textbf{18}, 4831 (2001). F. Dahia and C. Romero, J. Math. Phys. \textbf{43}%
, 3097 (2002). E. Anderson, F. Dahia, James E. Lidsey, C. Romero, J.
Math.Phys. \textbf{44}, 5108 (2003). F. Dahia and C. Romero, Class. Quant.
Grav. \textbf{21}, \ 927 (2004). F. Dahia and C. Romero, Class. Quant. Grav. 
\textbf{22}, 5005 (2005)

\bibitem{kaluza} T. Kaluza, \textit{Sitz. Preuss. Akad. Wiss. }\textbf{33},
966 (1921). O. Klein, \textit{Z. Phys}. \textbf{37}, 895 (1926). T.
Appelquist, A. Chodos and P. Freund,\textit{\ Modern Kaluza-Klein Theories},
Addison-Wesley, Menlo Park, 1987.

\bibitem{Ponce} P. S. Wesson and J. Ponce de Leon, J.Math.Phys.\textbf{33},
3883 (1992).

\bibitem{maia} Maia, M. D., \textit{Hypersurfaces of five-dimensional vacuum
space-times}, gr-qc/9512002.

\bibitem{Ponce2} J. Ponce de Leon, Gen. Rel. Grav. \textbf{20}, 539.

\bibitem{Weyl} H. Weyl, Sitzungesber Deutsch. Akad. Wiss. Berlin, 465
(1918). H. Weyl, \textit{Space, Time, Matter} (Dover, New York, 1952)

\bibitem{Pauli} See, for instance, W. Pauli, \textit{Theory of Relativity}
(Dover, New York, 1981). See, also, L. O'Raiefeartaigh and N. Straumann,
Rev. Mod. Phys. \textbf{72}, 1 (2000).

\bibitem{we1} F. Dahia, G. A. T. Gomez and C. Romero, \
[ArXiv:gr-qc/0711.2754].

\bibitem{Novello} M. Novello, L.A.R. Oliveira, J.M. Salim, E. Elbas, Int. J.
Mod. Phys. \textbf{D1} (1993) 641-677.\ J. M. Salim and S. L. Saut\'{u},
Class. Quant. Grav. \textbf{13}, 353 (1996). H. P. de Oliveira, J. M. Salim
and S. L. Saut\'{u}, Class.Quant.Grav. \textbf{14},\textbf{\ }2833 (1997).
V. Melnikov, \textit{Classical Solutions in Multidimensional Cosmology} in
Proceedings of the VIII Brazilian School of Cosmology and Gravitation II
(1995), edited by M. Novello (Editions Fronti\`{e}res) pp. 542-560, ISBN
2-86332-192-7.

\bibitem{Israelit} Mark Israelit, Found. Phy. Vol. 35 N%
${{}^\circ}$
10 (2005) 1725-1748.

\bibitem{Dahia1} F. Dahia, L. F. P. da Silva, C. Romero e R. Tavakol, J.
Math. Phys. \textbf{48}, 072501 (2007). F. Dahia, L. F. P. da Silva, C.
Romero e R. Tavakol, Gen. Rel. Grav. \textbf{40}, 1341 (2008).

\bibitem{Wheeler} See, for instance, C. W. Misner, K. S. Thorne and J. A.
Wheeler, \textit{Gravitation}, Ch. 21 (Freeman, 1973)
\end{thebibliography}
\end{document}